\documentclass[a4paper,fleqn]{cas-dc}
\let\printorcid\relax 

\usepackage[authoryear,longnamesfirst]{natbib}
\usepackage{verbatim}
\usepackage{graphicx}
\usepackage{url}
\usepackage{import}
\usepackage{subcaption}
\usepackage{multirow}
\usepackage{color, colortbl}
\definecolor{Gray}{gray}{0.9}
\usepackage{color,soul}
\usepackage{tabularray}
\usepackage{breakurl}
\usepackage{nomencl}
\usepackage{framed}

\begin{document}
\let\WriteBookmarks\relax
\def\floatpagepagefraction{1}
\def\textpagefraction{.001}
\shorttitle\relax
\shortauthors{Aharon et~al.}  
\title [mode = title]{A Classification-by-Retrieval Framework for Few-Shot Anomaly Detection to Detect API Injection Attacks} 
\author[1,2]{Udi Aharon}
\cormark[1]
\author[1,2]{Ran Dubin}
\author[2,3]{Amit Dvir}
\author[2,3,4]{Chen Hajaj}

\cortext[1]{Corresponding author at: Department of Computer Science, Ariel University, Golan Heights 1, 4077625, Ariel, Israel. E-mail addresses: udi.aharon@msmail.ariel.ac.il (U. Aharon), rand@ariel.ac.il (R. Dubin), amitdv@ariel.ac.il (A. Dvir), chenha@ariel.ac.il (C. Hajaj).}

\affiliation[1]{organization={School of Computer Science, Ariel University},
            addressline={Golan Heights 1}, 
            city={Ariel},
            postcode={4077625}, 
            country={Israel}}
            
\affiliation[2]{organization={Ariel Cyber Innovation Center, Ariel University},
            addressline={Golan Heights 1}, 
            city={Ariel},
            postcode={4077625}, 
            country={Israel}}

\affiliation[3]{organization={Data Science and Artificial Intelligence Research Center, Ariel University},
            addressline={Golan Heights 1}, 
            city={Ariel},
            postcode={4077625}, 
            country={Israel}}

\affiliation[4]{organization={Department of Industrial Engineering \& Management, Ariel University},
            addressline={Golan Heights 1}, 
            city={Ariel},
            postcode={4077625}, 
            country={Israel}}



\begin{abstract}
Application Programming Interface (API) Injection attacks refer to the unauthorized or malicious use of APIs, which are often exploited to gain access to sensitive data or manipulate online systems for illicit purposes. Identifying actors that deceitfully utilize an API poses a demanding problem. Although there have been notable advancements and contributions in the field of API security, there remains a significant challenge when dealing with attackers who use novel approaches that don't match the well-known payloads commonly seen in attacks. Also, attackers may exploit standard functionalities unconventionally and with objectives surpassing their intended boundaries. Thus, API security needs to be more sophisticated and dynamic than ever, with advanced computational intelligence methods, such as machine learning models that can quickly identify and respond to abnormal behavior. In response to these challenges, we propose a novel unsupervised few-shot anomaly detection framework composed of two main parts: First, we train a dedicated generic language model for API based on FastText embedding. Next, we use Approximate Nearest Neighbor search in a classification-by-retrieval approach. Our framework allows for training a fast, lightweight classification model using only a few examples of normal API requests. We evaluated the performance of our framework using the CSIC 2010 and ATRDF 2023 datasets. The results demonstrate that our framework improves API attack detection accuracy compared to the state-of-the-art (SOTA) unsupervised anomaly detection baselines.
\end{abstract}

\begin{highlights}
\item A novel unsupervised few-shot anomaly detection framework to detect API Injection attacks.
\item Introduced a tokenizer designed to capture and emphasize language factors specific to APIs, addressing unique NLP challenges.
\item Leveraged FastText embedding combined with Approximate Nearest Neighbor search, employing a Classification-by-retrieval approach.
\item Validated the framework using public HTTP datasets against cutting-edge techniques.
\end{highlights}


\maketitle

\section{Introduction}
Application Programming Interface (API) refers to a set of routines, procedures, resources, and protocols that permit the interaction between software systems and data exchange services ~\citep{balsari2018reimagining, ofoeda2019application, mendoza2018mobile}. APIs are an evolving technology for orchestrating applications utilizing web technology~\citep{benzaid2020zsm, ibm2016innovation}. Recently, it has been argued that we live in an API economy~\citep{ibm2016innovation} due to the growing interconnectedness of people, applications, and systems, all of which are powered by APIs. These interfaces now serve as the foundational framework of the digital ecosystem, establishing connections between industries and economies to foster value creation and cultivate innovative capabilities~\citep{ofoeda2019application}. APIs find utilization across a broad spectrum of services, including Web applications, Operation Systems (OS), Databases, and Hardware~\citep{reddy2011api}. The increasing ease of building web applications has led to a rise in agile development, where even inexperienced engineers can deploy applications. However, this approach often lacks strong security design or hardening planning. This can lead to vulnerabilities in application logic and inadequate consideration of security impacts. For instance, failure to properly constrain resources or access levels could lead to denial of service attacks~\citep{sun2022research}. As a result, the extensive adoption of web APIs has heightened the potential of user safety and privacy breaches, making APIs a prime target for cyber attackers~\citep{mendoza2018mobile}. In recent years, there have been several high-profile API attacks~\citep{forbeszoom2020,bbccap12019,therecordhilton2023,thenytimesequifax2017}, such as the Zoom video conferencing platform in 2020. 

To address these challenges, the Open Web Application Security Project (OWASP\footnote{\url{https://owasp.org/}}) provides resources, tools, and best practices to help organizations and developers enhance the security of their web applications and protect against malicious attacks. One of the most well-known contributions is the OWASP API Top 10, which outlines the ten most critical API security risks~\citep{fett2016comprehensive}. Injection flaws enable attackers to inject malicious code through an application into another system. These flaws encompass a variety of attacks, such as Directory Traversal, SQL Injection (SQLi), LDAP Injection, and Cross-Site Scripting (XSS). OWASP classifies Injection flaws under the broader category of Unsafe Consumption of APIs security risks\footnote{\url{https://owasp.org/API-Security/editions/2023/en/0xaa-unsafe-consumption-of-apis/}}.

Despite the progress made in prior research on utilizing machine learning and deep learning models for protecting APIs against both known and unknown attacks~\citep{chan2023transformer,baye2021api,harlicaj2021anomaly}, there are lingering concerns that remain unresolved. These concerns encompass several challenges, such as effectively detecting zero-day vulnerabilities, minimizing false positives, and addressing real-time and continuous protection requirements. Given that a zero-day API attack involves an unknown vulnerability that the security solutions, such as web application firewalls, are unaware of, it becomes necessary to employ few-shot learning techniques. The anomaly detection model needs to leverage its comprehension of previously encountered samples to make predictions on new, unseen samples. In light of the requirement to improve traditional methods for training binary anomaly classifiers, the utilization of Classification-by-Retrieval presents a solution~\citep{shen2017classification}. This approach enables the construction of neural network-based classifiers without requiring computationally intensive training procedures. Consequently, it facilitates the development of a lightweight model that can be trained with minimal examples per class or even a model capable of classifying multiple classes~\citep{shi2022nearest,qamar2008similarity,valero2023multilabel}. In this context, we propose a novel unsupervised few-shot anomaly detection framework, utilizing FastText embedding and Approximate Nearest Neighbor search (FT-ANN). This framework offers a unique representation for the API request, facilitating intelligent segmentation based on the API endpoint. The term "endpoint" refers to any data or metadata that may represent a type, an origin, or identification of a respective API request. Endpoint is defined as the combination of the method, host, and path~\citep{battle2008bridging}. This paper’s contributions are itemized as follows: 
\begin{enumerate}[\textbullet]
\item We introduce a novel unsupervised few-shot anomaly detection framework, utilizing FastText embedding and ANN search, which leverages a Classification-by-retrieval approach. This framework employs a Classification-by-retrieval approach and allows for training a single retrieval model that can handle multiple baselines concurrently, such as API endpoints or multiple API domains. Our approach provides a more efficient and scalable solution for anomaly detection by reducing the number of required models and supporting incremental index updates. This is in contrast to traditional models, which usually require the entire dataset to be present for training new examples. Furthermore, the high performance and speed of the in-memory similarity search, widely adopted in many ecosystems due to its flexible nature, significantly enhance the overall efficiency of our framework.
\item We define a novel tokenizer that specifically emphasizes the language factors present in APIs, addressing the unique challenges associated with API-based natural language processing. APIs are defined by a URL-based syntax in which each URL corresponds to a particular resource or action, and they include fundamental actions such as GET and POST, determining the structure of requests and responses. Additionally, APIs utilize the standard structure of HTTP headers to transmit metadata pertaining to both the request and the response. Unlike existing tokenizers, our approach considers these specific language characteristics of APIs, enabling more accurate and efficient processing of API-related text.
\item Our language model is designed to be domain-agnostic, eliminating the need for retraining when transitioning to different API domains. This flexibility allows our model to seamlessly serve various domains without sacrificing performance or requiring additional training efforts. 
\item Our model's agnostic nature allows it to seamlessly adapt and address the unique requirements and challenges, as also outlined in the OWASP Top 10 API vulnerabilities, posed by different API forms, including REST, GraphQL, gRPC, and WebSockets, regardless of the emphasis on HTTP datasets during the demonstration.
\end{enumerate}

The remaining sections of this paper are structured in the following manner. In Section~\ref{sec:related_work} we provide an overview of the relevant literature; The architecture of the FT-ANN framework is outlined in Section~\ref{sec:architecture}; Section~\ref{sec:datasets} describes the datasets that were used for this paper; The experimental design is outlined in Section~\ref{sec:evaluation_metrics}; Section~\ref{sec:results} outlines the experimental results, which focus on measuring the effectiveness the FT-ANN framework and presenting experimental results compared to various state-of-the-art benchmarks; Finally, we draw conclusions, current research limitations, and suggestions for future work in Section~\ref{sec:conclusions}. For ease of reading, we provide a list of common abbreviations in Table~\ref{tab:abbreviations}.

\begin{table}[]
\resizebox{\columnwidth}{!}{%
\begin{tabular}{ll}
        \hline
        \textbf{Abbreviation} & \textbf{Definition} \\
        \hline
            AE & Autoencoder \\
            ANN & Approximate Nearest Neighbor \\
            API & Application Programming Interface \\
            CBLOF & Clustering-Based Local Outlier Factor \\
            CNN & Convolutional Neural Network \\
            DeepSVDD & Deep Support Vector Data Description \\
            FB & Feature Bagging \\
            GRU & Gated Recurrent Unit \\
            HBOS & Histogram-based Outlier Detection \\
            IF & Isolation Forest \\
            KDE & Kernel Density Estimation \\
            LMDD & Linear Model Deviation-based Outlier Detection \\
            LOF & Local Outlier Factor \\
            LSTM & Long Short-Term Memory \\
            MCD & Minimum Covariance Determinant \\
            NLP & Natural Language Processing \\
            OCSVM & One-class SVM \\
            PCA & Principal Component Analysis \\
            QPS & Queries per second \\
            RDA & Regularized Deep Autoencoder \\
            RNN & Recurrent Neural Networks \\
            SOTA & State-of-the-art \\
            SQLi & SQL Injection \\
            SVM & Support Vector Machine \\
            XSS & Cross-Site Scripting \\
\end{tabular}%
}
\caption{List of Abbreviations.}
\label{tab:abbreviations}
\end{table}

\section{Related Work}
\label{sec:related_work}
This section reviews relevant research on API and web traffic classification, focusing on the use of machine learning, deep learning, and natural language processing (NLP) techniques.

~\cite{9376034} proposed supervised sequence models based on Recurrent Neural Networks (RNN) to identify malicious injections in API requests in addition to a heuristic rule that classified 10\% of each request sequence with a probability of 60\% as valid to minimize the number of false positives. However, such rules can introduce subjective biases and limitations. They generated a custom-labeled dataset but used only the request payloads. This presents a significant constraint in real-world scenarios, where malicious actors could potentially manipulate a majority of the request components, including headers~\citep{ombagi2017time}. They compared six unidirectional and bidirectional RNN models by evaluating various performance measures and showed a 50\% decrease in false positive cases.~\cite{JEMAL202158} suggested the Convolutional Neural Network (CNN) method to detect web attacks. They concluded that appropriately adjusting hyper-parameters and employing a data pre-processing approach significantly impacts the detection rate.~\cite{jemal2020m} propose a supervised Memory CNN that combines a CNN and Long Short-Term Memory (LSTM) to identify patterns of malicious requests within sequences of requests.~\cite{niu2020high} propose a technique for detecting web attacks based on a supervised CNN and Gated Recurrent Unit (GRU). They extracted statistical features and utilized a Word2Vec model to extract word embeddings, resulting in a 3-dimensional input for the suggested CNN-GRU method. A fully connected layer was used for classification.~\cite{yu2020detecting} combined a CNN and Support Vector Machine (SVM) to detect malicious web server requests.~\cite{baye2021api} also utilized SVM with a Linear Kernel as a two-class classifier to identify anomalous API requests. To form a training dataset that accurately represented authentic API logs, they employed a technique for outlier detection based on Gaussian Distribution, generating a synthetic dataset with labeled examples.

~\cite{Gniewkowski2021HTTP2vecEO} suggested an NLP-based self-supervised anomaly detection methodology that employs RoBERTa for embedding HTTP requests. A notable limitation in their method is the relatively long training and inference time, which presents a considerable drawback in real-time systems. ~\cite{moradi2019auto} introduced an unsupervised anomaly detection technique utilizing Autoencoder (AE) and LSTM for feature extraction and Isolation Forest (IF) for classification. However, they faced limitations related to the choice of encoding method for HTTP data, the non-stationary nature of HTTP data, and the integration of multiple feature sets, which may impact the effectiveness of the proposed approach. In a more recent study~\citep{moradi2021auto}, an unsupervised Deep Support Vector Data Description (DeepSVDD) method was proposed. They conducted a comparison between two feature extraction approaches, namely bigram (2-gram) and one-hot. They pointed out the limitation of lacking support for data streams in incremental learning, emphasizing the need for future research to address this aspect.~\cite{mac2018detecting} also suggested employing AEs for unsupervised learning, evaluated various variants, and concluded that the Regularized Deep AE (RDA) performed the best. They processed the raw HTTP requests and tokenized the URLs by replacing the characters with their corresponding ASCII codes.~\cite{vartouni2018anomaly} used a Stacked AE to extract HTTP features and employed IF as the classifier. In a Stacked AE, the output of each AE serves as the input for the next layer. They demonstrated that using various activation functions and optimization techniques can improve the learning process.~\cite{demirel2023acum} proposed combining the Local Outlier Factor (LOF) algorithm and an AE to enhance classification accuracy. The study evaluates different combination approaches, including the Behavior Knowledge Space (BKS) for combining multiple classifiers based on their predictions and Weighted Borda Counts (wBorda) for weighted voting systems, to assess their effectiveness.

\renewcommand{\arraystretch}{1.5}
\begin{table*}[]
\centering
\begin{tabular}
{@{}p{4.3cm}p{1.8cm}lp{3.4cm}ll@{}}
\cline{1-6}
\multicolumn{1}{c}{\textbf{Ref}} &
  \multicolumn{1}{c}{\textbf{Task}} &
  \multicolumn{1}{c}{\textbf{Technique(s)}} &
  \multicolumn{1}{c}{\textbf{Method(s)}} &
  \multicolumn{1}{c}{\textbf{Dataset(s)}} &
   \\ \cline{1-6}
~\cite{yu2020detecting} &
  Supervised &
  TextCNN, SVM &
  Word embedding,t-SNE, Statistical features &
  CSIC 2010 &
   \\ \cline{1-6}
~\cite{niu2020high} &
  Supervised &
  CNN, GRU &
  Word embedding, Statistical features &
  CSIC 2010 &
   \\ \cline{1-6}
~\cite{jemal2020m} &
  Supervised &
  CNN, LSTM &
  ASCII embedding &
  CSIC 2010 &
   \\ \cline{1-6}
~\cite{9376034} &
  Supervised &
  RNN &
  Word embedding &
  Custom &
   \\ \cline{1-6}
~\cite{JEMAL202158} &
  Supervised &
  CNN &
  Word embedding, Character embedding &
  CSIC 2010 &
   \\ \cline{1-6}
\multirow{3}{*}{~\cite{Gniewkowski2021HTTP2vecEO}} &
  \multirow{3}{*}{Self-supervised} &
  \multirow{3}{*}{RoBERTa} &
  \multirow{3}{*}{BBPE, Doc2vec} &
  CSIC 2010 &
   \\ \cline{5-6}
 &
   &
   &
   &
  Custom &
   \\ \cline{5-6}
 &
   &
   &
   &
  CSE-CIC-IDS2018 &
   \\ \cline{1-6}
~\cite{mac2018detecting} &
  Unsupervised &
  AE &
  ASCII embedding &
  CSIC 2010 &
   \\ \cline{1-6}
~\cite{vartouni2018anomaly} &
  Unsupervised &
  SAE, IF &
  Character N-grams &
  CSIC 2010 &
   \\ \cline{1-6}
~\cite{moradi2019auto} &
  Unsupervised &
  AE LSTM &
  Character embedding &
  CSIC 2010 &
   \\ \cline{1-6}
\multirow{2}{*}{~\cite{moradi2021auto}} &
  \multirow{2}{*}{Unsupervised} &
  \multirow{2}{*}{DeepSVDD} &
  \multirow{2}{*}{Character embedding} &
  CSIC 2010 &
   \\ \cline{5-6}
 &
   &
   &
   &
  ECML/PKDD 2007 &

   \\ \cline{1-6}

\multirow{3}{*}{~\cite{demirel2023acum}} &
  \multirow{3}{*}{Unsupervised} &
  \multirow{3}{*}{AE, LOF} &
  \multirow{3}{*}{ASCII embedding} &
  CSIC 2010 &
   \\ \cline{5-6}
 &
   &
   &
   &
  Custom &
   \\ \cline{5-6}
 &
   &
   &
   &
  Fwaf &
   \\ \cline{1-6}

\multirow{2}{*}{FT-ANN} &
  \multirow{2}{*}{Unsupervised} &
  \multirow{2}{*}{Classification-by-retrieval} &
  \multirow{2}{*}{Sentence embedding} &
  CSIC 2010 &
   \\ \cline{5-6}
 &
   &
   &
   &
  ATRDF 2023 &
   \\ \cline{1-6}
\end{tabular}%
    \caption{Overview of machine learning and NLP-based techniques for detecting attacks in HTTP traffic.}
    \label{tab:Summary_studies}
\end{table*}

\section{Framework}
\label{sec:architecture}
The analysis of API requests can be framed as a problem in NLP. One challenge lies in selecting a language model capable of generating a vector space representation. In our work, we decided to utilize the FastText~\citep{facebookai2017fastText} model. FastText embeddings primarily aim to factor in the internal structure of words rather than simply learning word representations. This feature is especially advantageous for morphologically complex languages, allowing representations for various morphological forms of words to be learned separately. FastText offers Skip-gram and Continuous Bag-of-Words (CBOW) models to compute word representations. Although CBOW learns faster than Skip-gram, Skip-gram outperforms CBOW on small datasets~\citep{bansal2018sentiment}. FastText adopts a character n-gram approach to tokenize words, which effectively tackles Out-of-vocabulary (OOV) problems. This method not only generates embeddings for common words but also for rare, misspelled, or previously unseen words in the training corpus. In the realm of API security, it is essential to examine the internal structure of requests to comprehend their purpose and context. Security threats might be hidden within seemingly innocuous text, which could include the use of uncommon or previously unseen terms, or even efforts to mask their intention through spelling errors~\citep{xiao2016sql}. Moreover, APIs might be required to accommodate a diverse range of languages, including those characterized by intricate morphological structures~\citep{seyyar2022attack}.

Prior works have proposed various methods for detecting anomalies. While most of these approaches focus on classic classification tasks, we propose utilizing the ANN vector similarity method for identifying in-distribution records. The similarity search concept pertains to the process of identifying data points in a dataset that demonstrate similarities with a specific pattern, commonly referred to as a query~\citep{chavez2001searching}. To measure the similarity of a pair of data points, a distance function is used, where a small distance indicates that the two points are more similar or "closer" to each other~\citep{ponomarenko2014comparative}. The Nearest Neighbor (NN) search is a specific type of similarity search used to identify data points nearest in distance to a provided query point~\citep{indyk1998approximate}. The ANN allows search despite the possibility of not retrieving all neighbors in a metric space~\citep{indyk1998approximate}.

While both ANN and traditional NN are rooted in the fundamental concept of similarity search, traditional NN search involves an exhaustive examination of all data points to find the NN, which can be prohibitively time-consuming for large datasets~\citep{indyk1998approximate}. In contrast, ANN employs techniques that trade-off a slight loss in precision for substantial gains in speed, enabling the identification of ANN without examining every data point. This efficiency becomes especially crucial in real-world API applications where rapid response times are essential. Additionally, ANN methods are adaptable to high-dimensional spaces~\citep{chavez2001searching,hajebi2011fast}, where traditional NN searches can suffer from the "curse of dimensionality", in which the computational requirements for exact NN search become prohibitively high as the dimensionality of the dataset increases~\citep{indyk1998approximate}.

\begin{figure*}[!htb]
\centering
\includegraphics[width=5in]{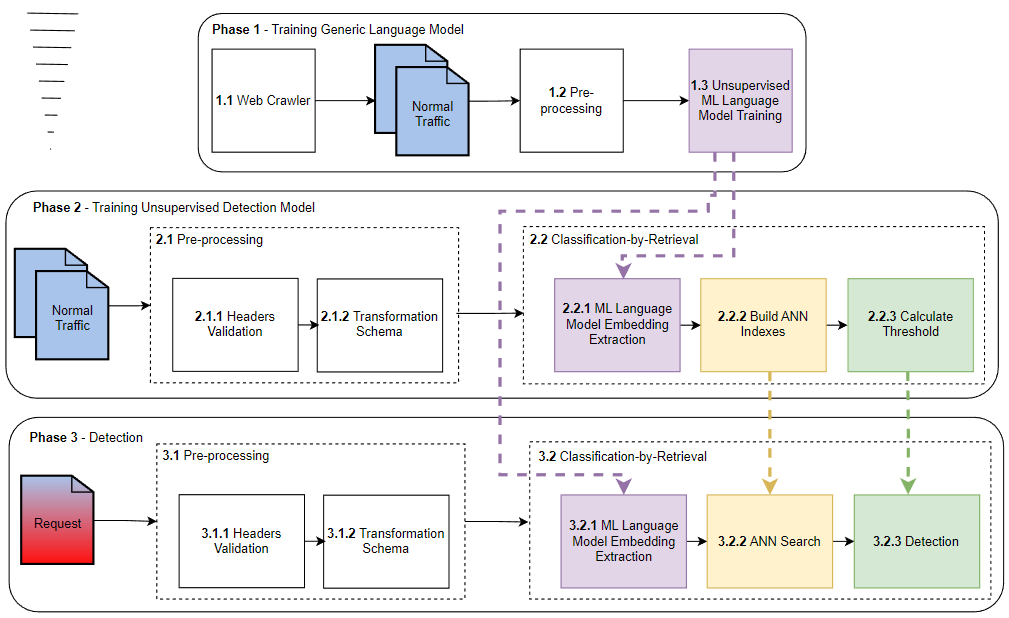}
\caption{The FT-ANN Framework: Training and Inference Methodology.}
\label{fig:ann-ft}
\end{figure*}

The framework proposed (depicted in Figure~\ref{fig:ann-ft}) comprises a combination of a FastText embedding network and a retrieval layer, which includes an ANN matching component and a result aggregation component built upon it, forming the FT-ANN system. In Phase 1 (Figure~\ref{fig:ann-ft}), we initiate the process by training a generic language model to serve as a reliable baseline for the detection model. To gather a substantial dataset, we crawled a random sample of websites from the Tranco top websites list, collecting over a million examples of normal web traffic. This language model now serves as a pre-trained baseline, applicable to any real-world API anomaly detection system. 

Proceeding to Phase 2 (Figure~\ref{fig:ann-ft}), the unsupervised detection model is trained, consisting of a pre-processing step and a Classification-by-retrieval framework. The pre-processing step validates the HTTP headers against the standard structure. Additionally, a unique data transformation is applied to simplify the API vocabulary. The detection model is trained solely on normal API traffic and constructs a single ANN model for all endpoints. For each endpoint, a threshold is calculated and utilized during the detection stage. 

Finally, in Phase 3 (Figure~\ref{fig:ann-ft}), the model's performance is validated by allowing an index search for every request. During validation, we standardize the request using the same pre-processing stages applied during the detection model training phase. The text is then transformed into a vector representation using our generic language model through inference. These embedding vectors are used in the ANN search, which returns the K-NNs from the specific endpoint index collection, followed by a maximum distance scaling layer. The final score obtained enables the model to ascertain whether the incoming API request is normal or an anomaly by comparing it with the pre-defined threshold for each API endpoint.

\subsection{Data Pre-Processing}
\label{sec:preprocess}
For both training and inferring the detection model, we employ a unique pre-processing technique consisting of three phases, as depicted in Figure~\ref{fig:ann-ft} in steps 2.1 and 3.1. First, decoding URL special characters, decompressing request body content, and converting every character within the request data string into lowercase. Then, validating request headers to ensure they are formatted correctly and extracting the endpoint definition, as depicted in Figure~\ref{fig:ann-ft} 2.1.1 and 3.1.1. API request headers typically provide information about the request context, supply authentication credentials, and provide information about the client (e.g., a person, a computing device, and/or a browser application) that had initiated the API ~\citep{buchanan2018analysis}. API request header fields are typically derived from a limited set of options. Accordingly, the data preparation also uses a fixed set of rules to validate the content of request headers and filter-out headers according to these rules. Headers that include valid or approved strings may be transferred to their destination as an API request and may be excluded from additional processing. For example, request headers may include host strings, which specify host or Internet Protocol (IP) addresses and/or port numbers of a server to which the API request is being sent. Valid IPv4 syntax should have the following format: $(0 \leq n < 256).(0 \leq n < 256).(0 \leq n < 256).(0 \leq n < 256):(1 \leq n \leq 65535)$.

Lastly, as shown in Figure~\ref{fig:ann-ft} 2.1.2 and 3.1.2, we convert received requests into abstracted versions based on a conversion schema. For instance, we replace non-numeric single characters with the string "chr", which serves as a representative, abstract version of the original request string. Another example involves converting non-textual symbols into pre-defined textual strings. For instance, colons (":") may be converted to the string "colon".
During the pre-processing stage, the API language has been refined to achieve optimally, enabling it to describe various transactions in a consistent format without sacrificing their original meanings. In fact, in most cases, these requests have been intelligently merged into a single text representation. This consolidation not only streamlines the data but also ensures that the essential information pertaining to different transactions remains preserved.

\begin{figure}[!htb]
\centering
\includegraphics[width=3.5in]{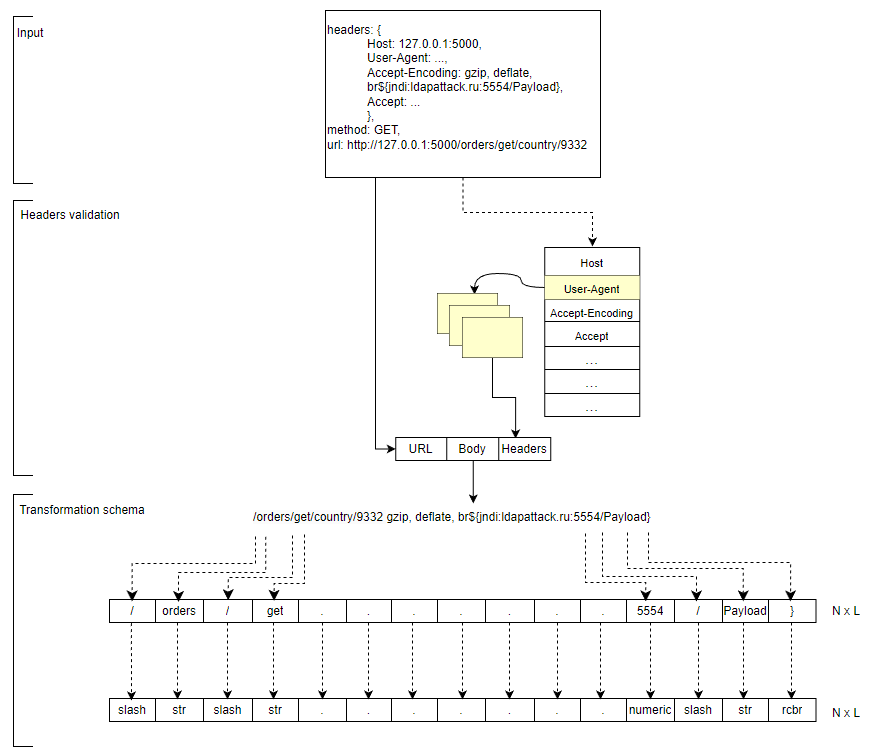}
\caption{Data pre-processing steps}
\label{fig:data_pre_processing}
\end{figure}

\subsection{Unsupervised ML Language Model}
\label{sec:unsupervised_ml}
Our framework leverages FastText for unsupervised learning. FastText possesses the capability to encapsulate substantive knowledge about words while integrating morphology details, a crucial aspect of API attack detection. While deep learning models have excelled and commonly provide state-of-the-art results across various NLP tasks, to the best of our knowledge, no previous NLP pre-trained model on the API traffic domain has been publicly published. In the training phase of the method, we built a single generic FastText language model (based on  ~\citep{facebookai2017fastText}) from scratch using the normal API traffic collected by crawling Tranco's list of the most popular websites, as can be seen in Figure~\ref{fig:ann-ft} Phase 1. For training the generic language model, we used the default hyper-parameters of  ~\citep{facebookai2017fastText}, which encompass a learning rate of 5\%, a word vector size of 100, and a context window size of 5. The model was then trained for 5 epochs. We utilized the CBOW model as the dataset is relatively large, and CBOW embeddings are precise enough for anomaly detection and computed in a shorter time than skip-gram~\citep{joulin2016bag}. For training the detection model and inferencing, we extract the vector representation of words for every input line.

\subsection{ANN}
\label{sec:ann}
We obtained ANN to identify the normal representation of an API endpoint. We train a single detection model, which is used to describe a normal representation of all endpoints. During the detection model training stage, each API request is represented as a vector in the textual embedding space, including endpoint information.
We employ cosine distance to measure the similarity between data points as it has been applied in numerous text mining endeavors, such as text classification, and information retrieval ~\citep{li2013distance}. Additionally, it has been proven to be effective for Out-of-Distribution (OOD) detection tasks ~\citep{techapanurak2020hyperparameter,hsu2020generalized}. Cosine similarity is a widely used measure of similarity that calculates the angle formed by a pair of vectors. When measuring the similarity between two patterns, the Euclidean distance increases as they become less similar, while the cosine similarity increases as they become more similar. Unlike Euclidean distance, cosine similarity is unaffected by the magnitude of the vectors being compared~\citep{xia2015learning}. 
The embedding vector feeds the Hierarchical Navigable Small Worlds (HNSW) graph~\citep{bilegsaikhan2014nmslib} to build new indexes of data points. During the detection stage, the model compares new API vectors of incoming API requests with API vectors of the same API endpoint information to evaluate the normality or anomaly of the incoming API request. The comparison is made by querying the similarity between the input vector to the closest k objects. 

The ANN search returns a set of IDs representing neighbors' points and the similarity score between the given point and its ID.
Max distance scaling is employed to scale the ANN similarity score within the given range. For every score, the maximum value gets transformed into a 0, and every other value is divided by the maximum similarity score in the range and then subtracted from 1. We use this method to invert the relationship between the original and the normalized scores to emphasize higher scores for smaller values. As in the cosine space, a smaller distance indicates that the two points are closer to each other. Let $X$ be the similarity score, the normalized score  X' is: $ X' = 1 - \left( \frac{X}{\max(X)} \right)$

\begin{figure}[!htb]
\centering
\includegraphics[width=3.4in]{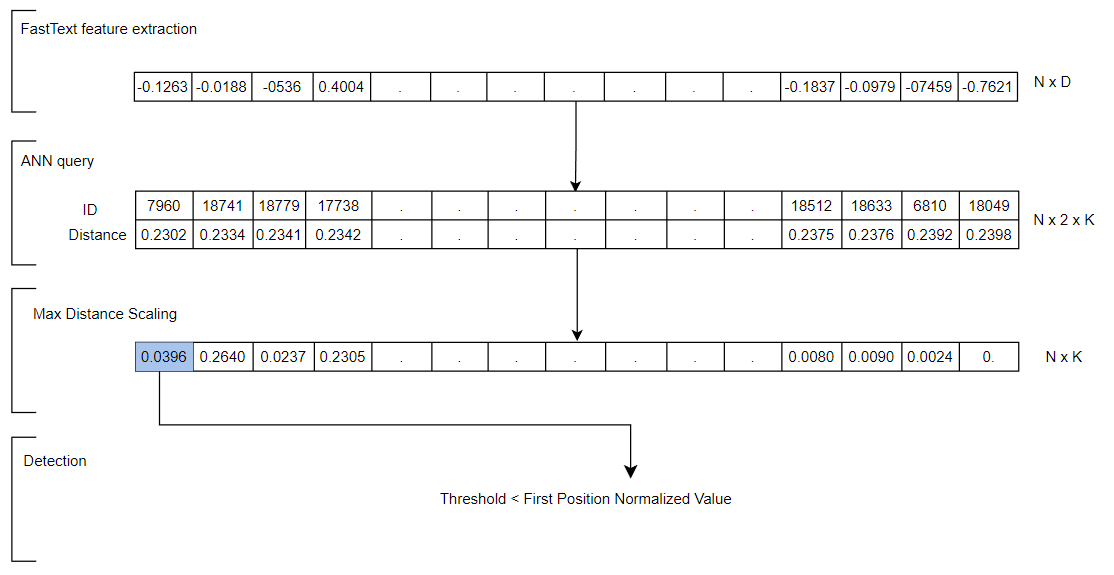}
\caption{Features extraction and classification-by-retrieval}
\label{fig:features_extraction_and_anomaly_detection}
\end{figure}

Lastly, we suggest an adaptive search for the best threshold for each API endpoint. As part of the detection model training, the model iteratively evaluates thresholds between 0 and 1, with increments of 0.1. The model considers the balance between precision and recall, as captured by the F1 score, to determine the optimal threshold value. As described in Figure ~\ref{fig:features_extraction_and_anomaly_detection}, in the detection stage, the first position score with the max normalized value is compared to the best threshold to determine anomaly.

\section{Experimental Design}
\label{sec:datasets}
API security papers are not as prevalent as the technology itself, despite being one of the most influential technologies~\citep{stone2022anari}. This is particularly evident in the scarcity of ready-to-use publicly available API datasets~\citep{torrano2011applying}. For this research, we used HTTP datasets which reflect REST API traffic. REST API interactions are typically carried out over HTTP, and the resulting HTTP requests and responses are logged by the web server~\citep{pautasso2010restful}.
Several of these datasets, such as ECML/PKDD 2007~\citep{ecml2007bee}, are outdated and unsuitable for usage, with some lacking traffic diversity and volume, and others failing to encompass a variety of attacks. Similarly, the CSE-CIC-IDS 2018~\citep{cse2018unb} dataset contains only a small portion of web attacks and directs normal traffic requests to various web applications. Additionally, it derives anomalous traffic from DVWA application routes, complicating endpoint isolation, and often lacks textual payloads, making it unsuitable for our research. Consequently, researchers resort to creating customized datasets by primarily utilizing open-source vulnerable web applications like DVWA, BWAPP, and Mutillidae, and employ automated penetration tools such as SQLMAP~\citep{damelesqlmap}, SQLNINJA~\citep{icesurfersqlninja}, and OWASP ZAP~\citep{bennetts2013owasp} to gather malicious payloads, including the Fwaf project~\citep{fwaf2017com}.

\begin{figure}[!htb]
  \centering
  \begin{subfigure}[b]{0.35\textwidth}  
    \centering
    \includegraphics[width=\textwidth]{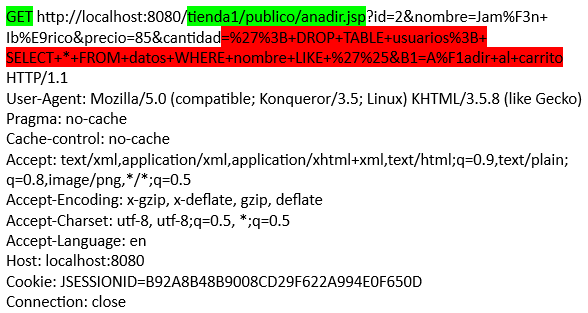}
    \caption{CSIC 2010 request sample}
    \label{fig:Request_CSIC_2010}
  \end{subfigure}%
  \hspace{0.02\textwidth}
    \begin{subfigure}[b]{0.35\textwidth}  
    \centering
    \includegraphics[width=\textwidth]{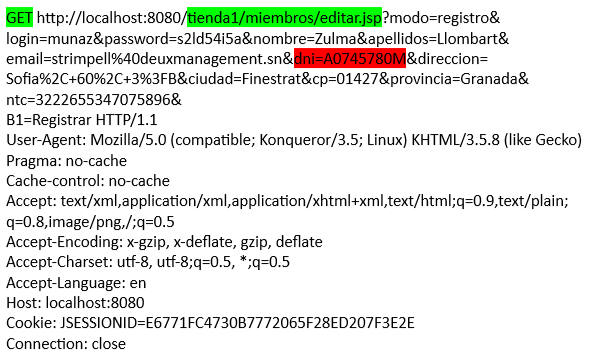}
    \caption{CSIC 2010 unintentional illegal request}
    \label{fig:csic_2010_unintentional_illegal_requests}
  \end{subfigure}%
  \hspace{0.02\textwidth}
  \begin{subfigure}[b]{0.35\textwidth}  
    \centering
    \includegraphics[width=\textwidth]{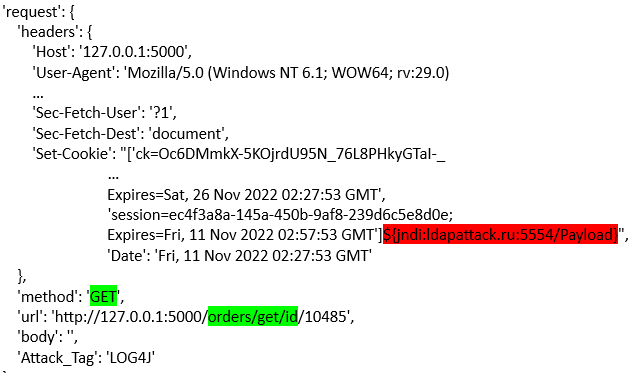}
    \caption{ATRDF 2023 request sample}
    \label{fig:request_atrdf_2023}
  \end{subfigure}
  \caption{An example of anomaly requests from CSIC 2010 and ATRDF 2023 datasets where endpoint definition is marked in green, malicious payload is marked in red.}
\end{figure}

Therefore, this research is evaluated on two datasets: CSIC 2010~\citep{gimenez2010http} and ATRDF 2023 ~\citep{atrdf2023lavian}. The HTTP CSIC 2010 dataset ~\citep{gimenez2010http} is widely used ~\citep{sanchez2021survey, seyyar2022attack, dawadi2023deep, Gniewkowski2021HTTP2vecEO, mac2018detecting, wang2018evaluating, moradi2019auto, jemal2020m, niu2020high, yu2020detecting, ito2018web,  jemal2021performance, yan2020web} in the field of malicious web traffic detection. This dataset was created by the Spanish Research National Council (CSIC). It is a sample of the traffic occurring on the Spanish e-commerce web application. The dataset includes attacks such as SQLi (Figure~\ref{fig:Request_CSIC_2010}), buffer overflow, information gathering, file disclosure, CRLF Injection, XSS, static attacks, and unintentional illegal requests. While unintentional illegal requests lack malicious intent, they deviate from the typical behavior of the web application and exhibit a different structure compared to regular parameter values. For instance, as shown in Figure~\ref{fig:csic_2010_unintentional_illegal_requests}, an invalid DNI (Spanish national ID number) was marked as an anomaly.  
We divided the dataset into two segments: the training portion, which included 36,000 normal requests and was exclusively utilized for representation learning, and the inference portion, which comprised both 36,000 normal and 25,000 anomalous traffic that was encoded by the model and employed for detection. It has 38 different endpoints, 8 of which have no normal representation and were excluded from our experiment. 

The API Traffic Research Dataset Framework (ATRDF)~\citep{atrdf2023lavian} is a recently published HTTP dataset publicly available which includes 18 different API endpoints. The dataset includes attacks such as Directory Traversal, Cookie Injection, Log4j (Figure~\ref{fig:request_atrdf_2023}), RCE, Log Forging, SQLi, and XSS. The dataset contains 54,0000 normal and 78,000 abnormal sets of requests and responses. The training set represents 80\% of the normal samples, with the remaining 20\% and the abnormal samples used for testing.

In response to the unavailability of a publicly accessible pre-trained model specialized for the API domain, a generic language model was developed to establish a reliable baseline for various detection models. However, developing a robust language model necessitates a significant amount of training data. To address this requirement, an extensive dataset of 1,061,095 API examples was collected. This dataset was obtained by performing a comprehensive data collection process, involving the crawling of a random sample of websites from the Tranco top websites list\footnote{\url{https://tranco-list.eu/}}. The Tranco list is an invaluable resource for cyber-security research as it provides a publicly available compilation of the top one million most popular domains, ranked based on a combination of four reputable lists: Alexa, Cisco Umbrella, Majestic, and Quantcast. Additionally, the Tranco list offers the advantage of being able to filter out unavailable or malicious domains, making it a valuable asset for our research~\citep{pochat2018tranco}.

\section{Evaluation Metrics}
\label{sec:evaluation_metrics}
In the realm of anomaly detection, it is commonplace to integrate a binary classification layer into the model architecture. This is due to the fundamental objective of distinguishing between normal and abnormal instances. To evaluate our architecture, we suggest using several performance measures in various experiments, including precision, recall, accuracy, and F1-score. 
Actual values are represented as True and False by (1) and (0), respectively, and predicted as Positive and Negative values by (1) and (0), respectively. Predicted possibilities of classification models are obtained through the expressions TP, TN, FP, and FN.

\textbf{Precision}:
 evaluates the accuracy of the positive predictions using the ratio of correctly predicted positive instances out of all instances predicted as positive. It is obtained by dividing the number of true positives by the sum of true positives and false positives.
\begin{equation}
Precision = \frac{TP}{TP + FP}
\end{equation}

\textbf{Recall}:
 measures the proportion of actual positives identified correctly. It is obtained by dividing the number of true positives by the sum of true positives and false negatives.

\begin{equation}
Recall = \frac{TP}{TP + FN}
\end{equation}

\textbf{Accuracy}:
 measures the overall correctness of the model predictions. It is obtained by dividing the total number of correct predictions by the total number of predictions made by the model.

\begin{equation}
Accuracy = \frac{TP + TN}{TP + TN + FP + FN}
\end{equation}

\textbf{F1-score}:
 measures the accuracy of the instances that were classified incorrectly by a model.
It is obtained by taking the harmonic mean of precision and recall.

\begin{equation}
F1\text{-}score = 2 \times \frac{Precision \times Recall}{Precision + Recall}
\end{equation}

In the current problem, two classes are represented by Positive and Negative, where the positive class corresponds to an abnormal API request and the negative class corresponds to a normal API request. We also compared the training and inference times of the anomaly detection SOTAs and the queries per second (QPS) of the ANN algorithms.

\section{Experimental Results}
\label{sec:results}
\begin{figure}[!htb]
     \centering
     \begin{subfigure}[b]{0.35\textwidth}
         \centering
         \includegraphics[width=\textwidth]{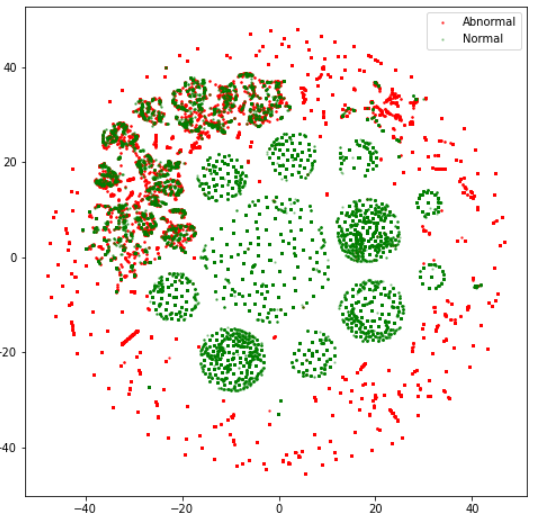}
         \caption{\textbf{CSIC 2010}}
         \label{fig:Tsne_CSIC_2010}
     \end{subfigure}
     \begin{subfigure}[b]{0.35\textwidth}
         \centering
         \includegraphics[width=\textwidth]{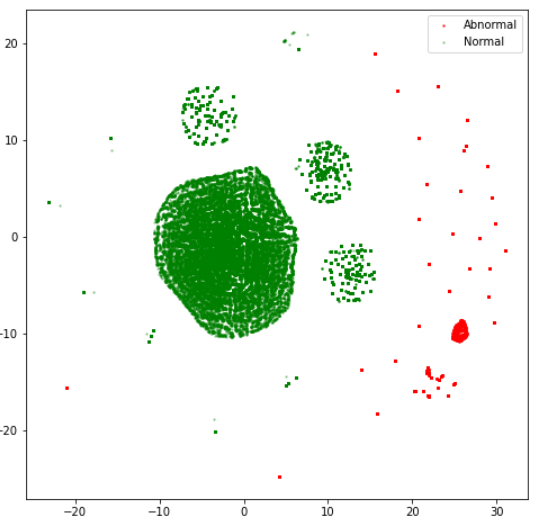}
         \caption{\textbf{ATRDF 2023}}
         \label{fig:Tsne_atrdf_2023}
     \end{subfigure}

        \caption{Visualization of FastText embedding using t-SNE shows the planar representation of the internal high-dimensional organization of the two classes. Each class is labeled with a unique color for clarity. Dimensionality reduction techniques, such as t-SNE, arrange data points so that those close together represent requests that FastText perceives as having similar patterns.}
        \label{fig:Tnse}
\end{figure}

To understand better the two datasets, we use the t-distributed Stochastic Neighbor Embedding (t-SNE) method for dimensionality reduction to graphically depict our high-dimensional datasets. The plots in Figure \ref{fig:Tnse} indicate distinct separation among classes within a reduced dimensional space. This suggests that requests with similarities tend to be grouped, enabling the exploration of a neighborhood for any given sample. Close records from different classes suggest that certain requests are quite similar, causing the embedding to overlap between the two classes. The result for ATRDF 2023 shows no overlap and clear separation between the classes while for CSIC 2010, some records from the two classes were found to be similar. We identified that most of the anomalous requests which overlap with normal requests have no malicious payload and are categorized as unintentional illegal requests.
					
Then, we compare our method with twelve detecting outlying objects in multivariate data baseline models~\citep{zhao2019pyod}: AE, Feature Bagging (FB), Histogram-based Outlier Detection (HBOS), IF, LOF, Minimum Covariance Determinant (MCD), One-class SVM (OCSVM), Principal Component Analysis (PCA), Deep One-Class Classification for outlier detection (DeepSVDD), Clustering-Based Local Outlier Factor (CBLOF), Kernel Density Estimation (KDE), Linear Model Deviation-based Outlier Detection (LMDD). We apply default hyper-parameters as provided by the original source code across all models for consistency and use the same pre-processed dataset to facilitate optimal comparison. 

Furthermore, we conducted an ANN benchmark to effectively contrast our framework with other nearest-neighbor algorithms. This benchmark involved generating algorithm instances based on a configuration file written in YAML format that defines the different methods and algorithms. At the top level, the point type is specified, followed by the distance metric, and finally, each algorithm implementation to be evaluated. Each implementation specifies the Python library, and additional entries provide the necessary arguments. For clarity, an illustrative example of this configuration file is presented in Figure~\ref{fig:Hnswlib-config}, while the complete file is available in the project GitHub repository~\citep{ftann2024aharon}. Both the "space" and "run\_groups" lists encompass arguments that should be included at the beginning of every invocation. Each algorithm defines one or more "run\_groups", each of which is expanded into several lists of arguments. The Cartesian product of these entries results in numerous argument lists. For instance, consider the Hnswlib entry depicted in Figure ~\ref{fig:Hnswlib-config}. This expands into three distinct algorithm instances: Cosine (Cosine Similarity), L2 (Squared L2), and IP (Inner product). Each of these instances undergoes training before being utilized for various experiments. Initially, experiments are conducted with different values of k, representing the number of neighbors to return (e.g., [10, 50, 100, 300, 400, 500, 1000, 2000, 2500, 3000]). Subsequently, experiments are conducted with varying ef\_construction values, which denote the size of the dynamic list used during index construction. A larger ef\_construction value indicates a higher quality index but also results in longer build times (e.g., [10, 20, 40, 80, 120, 200, 400, 600, 800]). Throughout each run, pertinent information is recorded, including the algorithm name, the time taken to construct the data structure used for indexing, and the outcomes of every query. These query outcomes encompass the neighboring points returned by the algorithm, the duration required to locate these neighbors, and the proximity between the neighbors and the query point. We leveraged the query results to compute the corresponding confusion matrix, enabling us to thoroughly evaluate the classification-by-retrieval performance of each algorithm. 
We conducted our benchmarking analysis by assessing the performance of various algorithm implementations, including Nmslib, Hnswlib, Bruteforce Blas, BallTree, KDtree, CKDtree, Annoy, Faiss, and RPForest, all of which were evaluated using the publicly accessible ANN-Benchmarks tool as a reference framework~\citep{aumuller2020ann}. 
To gain a deeper comprehension of how the embedding layer influences the detection outcome, we assessed all anomaly detection baseline models and the ANN benchmark using two additional prominent language models: BERT~\citep{devlin2018bert}, and RoBERTa~\citep{liu2019roberta}. Each model was individually trained from scratch and subsequently subjected to evaluation. We employed the RoBERTa model, specifically the RoBERTaForMaskedLM class, with a language modeling head on top\footnote{\url{https://huggingface.co/docs/transformers/model_doc/RoBERTa}}. This model was trained with a maximum sequence length of 512, utilizing 12 hidden layers and 12 attention heads. The training process spanned 10 epochs, employing a batch size of 16, which aligns with a similar approach in a prior study by ~\citep{Gniewkowski2021HTTP2vecEO}. Likewise, we employed the BERT model, specifically BertForMaskedLM, which also incorporates a language modeling head on top\footnote{\url{https://huggingface.co/docs/transformers/model_doc/bert}}. This model underwent training with a maximum sequence length of 512, utilizing 4 hidden layers and 4 attention heads, as was the case in a similar task outlined in ~\citep{guo2021logbert}.

\begin{figure}[htbp]
\centering
\begin{minipage}{\columnwidth}
\begin{verbatim}
  - name: Hnswlib
    library: Hnswlib
    method: [Hnswlib]
    space: [cosine,l2,ip]
    run_groups:
      K:
        query_args: [[10,50,100,300,400,
        500,1000,2000,2500,3000]]
      ef_construction:
        query_args: [[10, 20, 40, 80, 
        120, 200, 400, 600, 800]]
\end{verbatim}
\end{minipage}
\caption{Example configuration for the Hnswlib algorithm}
\label{fig:Hnswlib-config}
\end{figure}

We used Intel(R) Xeon(R) Silver 4214R CPU @ 2.40GHz to evaluate the effectiveness of each technique. Each endpoint was evaluated separately, and the average score for all endpoints was used to measure overall detector performance. We evaluate model performance by measuring execution time, inference time, and F1-score during training and testing. It is worth noting that detecting anomalies in real-time is critical in mitigating the effects of an attack. Therefore, it is advisable to analyze each phase separately to better understand each model's performance. Additionally, we aim to achieve better results compared to previous unsupervised SOTAs. Naturally, supervised-related works should not be compared since they rely on labeled data.

In the context of ANN search, selecting the optimal value for K involves determining the number of nearest neighbors to consider for predictions. The model performance evaluation and the choice of the optimized value of K rely on the F1 score, which provides a balanced assessment of precision and recall. By systematically iterating through a range of K values and assessing the F1 score for each value, we can identify the K value that yields the highest F1 score and use it in our final endpoint evaluation. During this experiment, the performance of the model was validated by iteratively testing different values of K, ranging from 1 to 1000.		

The results for the CSIC 2010 dataset, as can be seen from Table \ref{tab:CSIC-2010-performance}, comparing against the conventional outlier detection baselines, FT-ANN and PCA, demonstrate the shortest train times, with 0.0552 and 0.0575 seconds, respectively, whereas AE and LMDD are significantly slower. CBLOF has a relatively fast inference time but does not support incremental learning, requiring access to the entire dataset to fit the model. FT-ANN achieves the highest F1 score of 0.9713, indicating its balanced precision-recall performance. LMDD appears to exhibit the comparatively weakest performance across multiple metrics. This could be attributed to its reliance on a dissimilarity function that may not be well-suited to the complex and diverse anomalies present~\citep{arning1996linear}.

\renewcommand{\arraystretch}{1.5}
\begin{table*}[]
\resizebox{\textwidth}{!}{%
\begin{tabular}{llllllllllllllll}
\hline
 &
  \multicolumn{3}{c}{\textbf{F1 Score}} &
  \multicolumn{3}{c}{\textbf{Precision}} &
  \multicolumn{3}{c}{\textbf{Recall}} &
  \multicolumn{3}{c}{\textbf{Train Time (Sec)}} &
  \multicolumn{3}{c}{\textbf{Test Time (Sec)}} \\ \cline{2-16} 
\multirow{-2}{*}{} &
  \multicolumn{1}{l}{\textbf{BERT}} &
  \multicolumn{1}{l}{\textbf{FastText}} &
  \textbf{RoBERTa} &
  \multicolumn{1}{l}{\textbf{BERT}} &
  \multicolumn{1}{l}{\textbf{FastText}} &
  \textbf{RoBERTa} &
  \multicolumn{1}{l}{\textbf{BERT}} &
  \multicolumn{1}{l}{\textbf{FastText}} &
  \textbf{RoBERTa} &
  \multicolumn{1}{l}{\textbf{BERT}} &
  \multicolumn{1}{l}{\textbf{FastText}} &
  \textbf{RoBERTa} &
  \multicolumn{1}{l}{\textbf{BERT}} &
  \multicolumn{1}{l}{\textbf{FastText}} &
  \textbf{RoBERTa} \\ \hline
{\textbf{FT-ANN}} &
  \multicolumn{1}{l}{{0.9675}} &
  \multicolumn{1}{l}{{\textbf{0.9713}}} &
  {0.9664} &
  \multicolumn{1}{l}{{0.947}} &
  \multicolumn{1}{l}{{\textbf{0.9538}}} &
  {0.9465} &
  \multicolumn{1}{l}{{0.9982}} &
  \multicolumn{1}{l}{{\textbf{0.9954}}} &
  {0.9972} &
  \multicolumn{1}{l}{{0.4458}} &
  \multicolumn{1}{l}{{\textbf{0.0552}}} &
  {0.5626} &
  \multicolumn{1}{l}{{0.0353}} &
  \multicolumn{1}{l}{{\textbf{0.0075}}} &
  {0.0347} \\ \hline
\textbf{MCD} &
  \multicolumn{1}{l}{0.9622} &
  \multicolumn{1}{l}{0.9572} &
  0.9517 &
  \multicolumn{1}{l}{0.9377} &
  \multicolumn{1}{l}{0.9382} &
  0.9159 &
  \multicolumn{1}{l}{0.9971} &
  \multicolumn{1}{l}{0.9774} &
  0.9985 &
  \multicolumn{1}{l}{1269.2} &
  \multicolumn{1}{l}{6.3603} &
  1261.2 &
  \multicolumn{1}{l}{2.4071} &
  \multicolumn{1}{l}{0.0021} &
  2.5126 \\ \hline
\textbf{CBLOF} &
  \multicolumn{1}{l}{0.9404} &
  \multicolumn{1}{l}{0.9445} &
  0.9365 &
  \multicolumn{1}{l}{0.9297} &
  \multicolumn{1}{l}{0.9318} &
  0.9276 &
  \multicolumn{1}{l}{0.9531} &
  \multicolumn{1}{l}{0.9589} &
  0.9477 &
  \multicolumn{1}{l}{2.8296} &
  \multicolumn{1}{l}{0.5747} &
  2.7095 &
  \multicolumn{1}{l}{0.0245} &
  \multicolumn{1}{l}{0.001} &
  0.0259 \\ \hline
\textbf{IF} &
  \multicolumn{1}{l}{0.9397} &
  \multicolumn{1}{l}{0.9419} &
  0.9362 &
  \multicolumn{1}{l}{0.9289} &
  \multicolumn{1}{l}{0.9323} &
  0.9288 &
  \multicolumn{1}{l}{0.9525} &
  \multicolumn{1}{l}{0.9538} &
  0.9464 &
  \multicolumn{1}{l}{16.066} &
  \multicolumn{1}{l}{0.5734} &
  15.979 &
  \multicolumn{1}{l}{0.3801} &
  \multicolumn{1}{l}{0.048} &
  0.4007 \\ \hline
\textbf{HBOS} &
  \multicolumn{1}{l}{0.9383} &
  \multicolumn{1}{l}{0.9393} &
  0.9367 &
  \multicolumn{1}{l}{0.9289} &
  \multicolumn{1}{l}{0.9311} &
  0.9274 &
  \multicolumn{1}{l}{0.9501} &
  \multicolumn{1}{l}{0.9502} &
  0.9483 &
  \multicolumn{1}{l}{0.9967} &
  \multicolumn{1}{l}{0.8365} &
  1.0126 &
  \multicolumn{1}{l}{0.0587} &
  \multicolumn{1}{l}{0.0159} &
  0.0613 \\ \hline
\textbf{PCA} &
  \multicolumn{1}{l}{0.938} &
  \multicolumn{1}{l}{0.9391} &
  0.9349 &
  \multicolumn{1}{l}{0.929} &
  \multicolumn{1}{l}{0.9304} &
  0.9271 &
  \multicolumn{1}{l}{0.9494} &
  \multicolumn{1}{l}{0.9504} &
  0.9454 &
  \multicolumn{1}{l}{27.829} &
  \multicolumn{1}{l}{0.0574} &
  28.412 &
  \multicolumn{1}{l}{2.635} &
  \multicolumn{1}{l}{0.0025} &
  2.7356 \\ \hline
\textbf{AE} &
  \multicolumn{1}{l}{0.938} &
  \multicolumn{1}{l}{0.9389} &
  0.9349 &
  \multicolumn{1}{l}{0.929} &
  \multicolumn{1}{l}{0.9303} &
  0.9271 &
  \multicolumn{1}{l}{0.9494} &
  \multicolumn{1}{l}{0.9502} &
  0.9454 &
  \multicolumn{1}{l}{82.872} &
  \multicolumn{1}{l}{75.0644} &
  84.402 &
  \multicolumn{1}{l}{0.0922} &
  \multicolumn{1}{l}{0.0668} &
  0.0978 \\ \hline
\textbf{KDE} &
  \multicolumn{1}{l}{0.937} &
  \multicolumn{1}{l}{0.9361} &
  0.9338 &
  \multicolumn{1}{l}{0.9283} &
  \multicolumn{1}{l}{0.9296} &
  0.9263 &
  \multicolumn{1}{l}{0.9483} &
  \multicolumn{1}{l}{0.946} &
  0.9441 &
  \multicolumn{1}{l}{86.800} &
  \multicolumn{1}{l}{3.671} &
  83.525 &
  \multicolumn{1}{l}{8.6993} &
  \multicolumn{1}{l}{0.3875} &
  8.9307 \\ \hline
\textbf{OCSVM} &
  \multicolumn{1}{l}{0.9368} &
  \multicolumn{1}{l}{0.936} &
  0.9326 &
  \multicolumn{1}{l}{0.9283} &
  \multicolumn{1}{l}{0.9295} &
  0.9255 &
  \multicolumn{1}{l}{0.948} &
  \multicolumn{1}{l}{0.9459} &
  0.9426 &
  \multicolumn{1}{l}{45.390} &
  \multicolumn{1}{l}{1.6392} &
  50.025 &
  \multicolumn{1}{l}{2.8314} &
  \multicolumn{1}{l}{0.0848} &
  3.4049 \\ \hline
\textbf{FB} &
  \multicolumn{1}{l}{0.9355} &
  \multicolumn{1}{l}{0.9354} &
  0.9319 &
  \multicolumn{1}{l}{0.9279} &
  \multicolumn{1}{l}{0.9303} &
  0.9255 &
  \multicolumn{1}{l}{0.946} &
  \multicolumn{1}{l}{0.9444} &
  0.9415 &
  \multicolumn{1}{l}{10.875} &
  \multicolumn{1}{l}{10.5476} &
  10.775 &
  \multicolumn{1}{l}{1.199} &
  \multicolumn{1}{l}{0.9365} &
  1.2491 \\ \hline
\textbf{LOF} &
  \multicolumn{1}{l}{0.9354} &
  \multicolumn{1}{l}{0.935} &
  0.9317 &
  \multicolumn{1}{l}{0.9279} &
  \multicolumn{1}{l}{0.9301} &
  0.9255 &
  \multicolumn{1}{l}{0.9459} &
  \multicolumn{1}{l}{0.9439} &
  0.9412 &
  \multicolumn{1}{l}{1.1774} &
  \multicolumn{1}{l}{1.0486} &
  1.1146 &
  \multicolumn{1}{l}{0.1291} &
  \multicolumn{1}{l}{0.0909} &
  0.1295 \\ \hline
\textbf{DeepSVDD} &
  \multicolumn{1}{l}{0.9187} &
  \multicolumn{1}{l}{0.9235} &
  0.8999 &
  \multicolumn{1}{l}{0.923} &
  \multicolumn{1}{l}{0.9381} &
  0.9155 &
  \multicolumn{1}{l}{0.921} &
  \multicolumn{1}{l}{0.9126} &
  0.8954 &
  \multicolumn{1}{l}{38.213} &
  \multicolumn{1}{l}{32.7192} &
  38.2421 &
  \multicolumn{1}{l}{0.0728} &
  \multicolumn{1}{l}{0.0486} &
  0.0679 \\ \hline  
\textbf{LMDD} &
  \multicolumn{1}{l}{0.3292} &
  \multicolumn{1}{l}{0.3017} &
  0.3101 &
  \multicolumn{1}{l}{0.5701} &
  \multicolumn{1}{l}{0.528} &
  0.5885 &
  \multicolumn{1}{l}{0.2476} &
  \multicolumn{1}{l}{0.2293} &
  0.225 &
  \multicolumn{1}{l}{11882} &
  \multicolumn{1}{l}{225.87} &
  11229 &
  \multicolumn{1}{l}{517.69} &
  \multicolumn{1}{l}{11.642} &
  471.22 \\ \hline
\end{tabular}%
}
\caption{Performance evaluations of FT-ANN on the CSIC 2010 dataset in comparison to baseline models for detecting outlying objects in multivariate data.}
\label{tab:CSIC-2010-performance}
\end{table*}

Several observations stand out when comparing the framework performance using BERT and RoBERTa embedding to FastText. Notably, BERT and RoBERTa result in longer training and testing times for all models. For instance, FT-ANN with FastText is approximately 87\%-90\% faster in training than RoBERTa and BERT, and similarly, around 78\% faster in testing. This increase in time could be attributed to the more computational complexity nature of BERT~\citep{yu2022u,xin2021berxit} and RoBERTa~\citep{ruckle2020adapterdrop} models. In terms of performance, when considering both RoBERTa and BERT embedding, FT-ANN continues to exhibit robust performance, achieving an F1-score of 0.9664 and 0.9675, respectively. This result suggests that the framework effectively leverages the contextual information embedded within the two language model representations to identify anomalies. The high precision and recall values for FT-ANN across all language models further support the framework's ability to maintain a fine balance between detecting true anomalies and minimizing false positives. MCD with RoBERTa embedding achieves a recall of 0.9985, indicating better performance in correctly identifying positive instances compared to FT-ANN with RoBERTa embedding, which achieved 0.9972. Another noteworthy observation lies in the performance of the FT-ANN framework when using RoBERTa embedding. Surprisingly, while RoBERTa is considered an even more advanced and powerful language model compared to BERT, the F1-score for FT-ANN drops slightly to 0.9664. This outcome raises questions about why the transition to RoBERTa, which typically exhibits superior performance across a range of natural language processing tasks~\citep{liu2019roberta,tarunesh2021trusting,rajapaksha2021bert}, did not lead to an improved performance for this specific outlier detection method. MCD, KDE, OCSVM, FB, LOF, and LMDD exhibit consistent or improved F1-scores with BERT. However, for DeepSVDD, the F1-scores drop slightly with BERT to 0.9187 and even more with RoBERTa to 0.8999.

\begin{figure*}[!htb]
\centering
\includegraphics[width=7in]{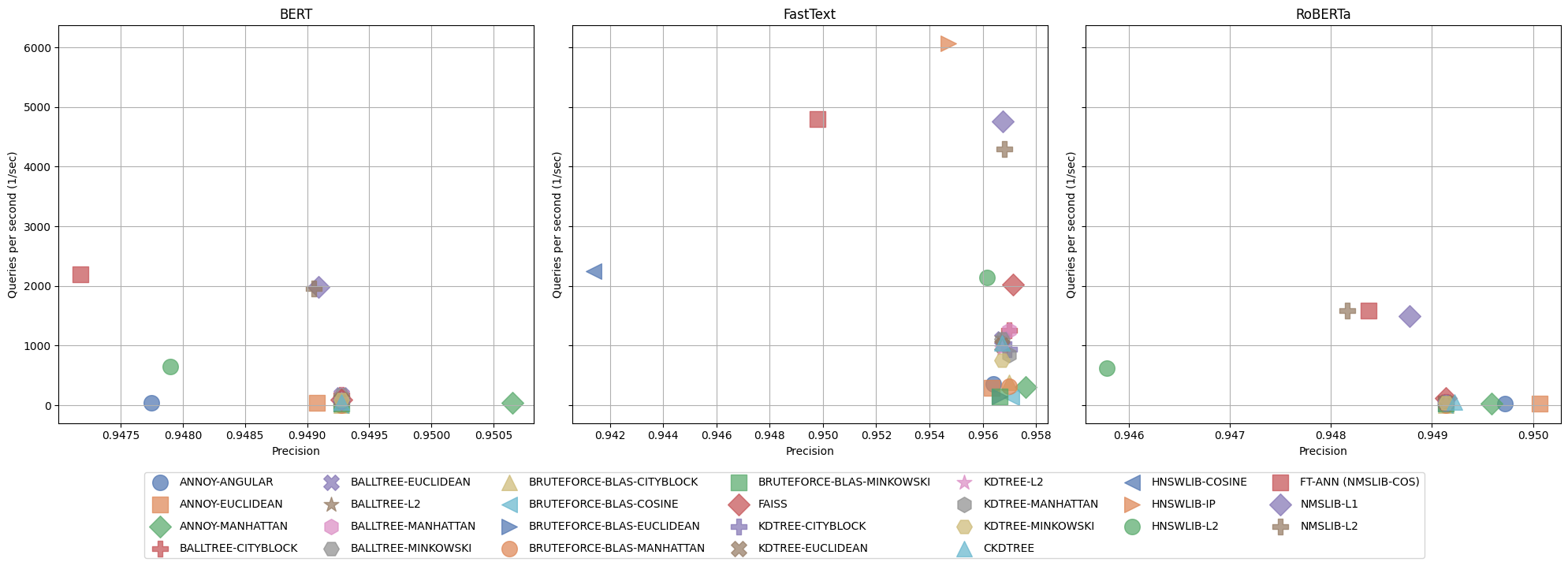}
\caption{Precision-QPS (queries per second) trade-off for top-performing in-memory ANN algorithms on the CSIC 2010 dataset — higher and to the right indicates better performance.}
\label{fig:csic_2010_ann_benchmark_precision_qps}
\end{figure*}

\begin{figure*}[!htb]
\centering
\includegraphics[width=7in]{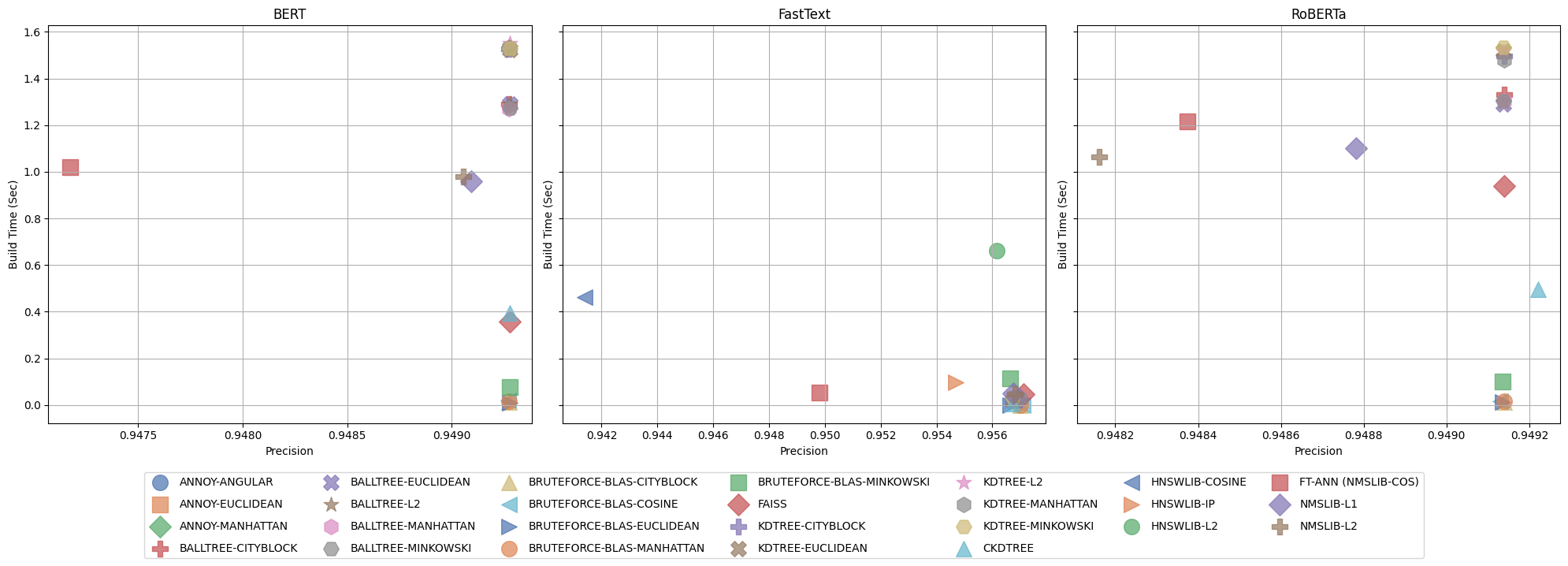}
\caption{Precision-Index build time trade-off in seconds for top-performing in-memory ANN algorithms on the CSIC 2010 dataset — lower and to the right indicates better performance.}
\label{fig:csic_2010_ann_benchmark_precision_build_time}
\end{figure*}

FT-ANN demonstrates superior performance over traditional outlier detection methods, while other ANN algorithms show similar performance with only minor variations. One of the most important challenges in any cybersecurity system is to minimize FP and take action in real time~\citep{kabir2018cyber}. Therefore, we conducted a trade-off analysis between QPS and index build time (training) with precision. In Figures~\ref{fig:csic_2010_ann_benchmark_precision_qps} and \ref{fig:csic_2010_ann_benchmark_precision_build_time}, we show only the leading algorithms for clarity, while the complete file is available in the project GitHub repository~\citep{ftann2024aharon}. The results show that FT-ANN (Nmslib-COS) offers a compelling balance of speed and accuracy, with a QPS of 4795.43 and a precision of 0.9498, FT-ANN excels in delivering fast query responses while maintaining competitive precision. Hnswlib-IP achieves the highest QPS (6,065.70) and slightly higher precision (0.9547), and Nmslib-L1 and Nmslib-L2 also show strong performance with high QPS and slightly higher precision. Algorithms like Bruteforce-Blas-Cosine and Faiss, despite their highest precision, significantly lag in QPS. When considering build time, FT-ANN demonstrates a moderate build time of 0.0519 seconds. While some algorithms like Bruteforce-Blas variants have significantly shorter build times (e.g., Bruteforce-Blas-Cityblock with 0.0004 seconds), they lag far behind in QPS. On the other hand, algorithms like Hnswlib-Cosine and Hnswlib-L2, despite achieving relatively high QPS, have longer build times (0.461 and 0.660 seconds, respectively). The Annoy algorithm's slower build time and lower QPS may be due to generating random projections and building binary trees for its search index, which requires significant computation to maintain data relationships within projection dimensions, leading to slightly longer training times compared to other models~\citep{cheng2021manifold}.

The results with BERT and RoBERTa embedding show significant differences in the performance metrics. FT-ANN with BERT achieves a precision of 0.9472, a QPS of 2,189.39, and a build time of 1.0195 seconds, while with RoBERTa, it attains a precision of 0.9484, a QPS of 1,591.54, and a build time of 1.2153 seconds. Other algorithms like Hnswlib-IP and Nmslib-L1 show decreased QPS and longer build times with BERT and RoBERTa embedding, reflecting the increased computational complexity and processing requirements of these advanced models. Despite maintaining high precision, algorithms such as Faiss and Bruteforce-Blas variants exhibit significantly lower QPS, highlighting their limitations in real-time applications. These insights underline that while BERT and RoBERTa embedding provide high precision, they often come at the cost of reduced speed and increased computational demands compared to FastText.

Despite not having the highest accuracy, Nmslib is used in industries such as Amazon Elasticsearch Service~\citep{aws2020build}. This indicates that although its accuracy might not be the absolute best compared to other ANN implementations, its simple deployment, ease of integration, and excellent balance between speed and accuracy align well with the requirements of real-world applications.

As seen in Table \ref{tab:csic_2010_sota}, FT-ANN demonstrates superior accuracy compared with current SOTA unsupervised learning models on the CSIC 2010 dataset, specifically AE variants~\citep{mac2018detecting, demirel2023acum, vartouni2018anomaly, moradi2019auto} and SVDD~\citep{moradi2021auto}.

The results for the ATRDF 2023 dataset, as can be seen from Table \ref{tab:ATRDF-2023-performance}, show that FT-ANN consistently achieves perfect scores across all metrics (F1 score, precision, recall) with BERT, FastText, and RoBERTa embedding. Comparatively, most of the baseline models also achieved high accuracy but failed to classify correctly all positive instances as reflected in the precision. DeepSVDD, AE, and IF perform well but exhibit longer build and test times. LMDD has higher build and test times, indicating again less efficient scalability.

\begin{figure}[!htb]
\centering
\includegraphics[width=0.5\textwidth]{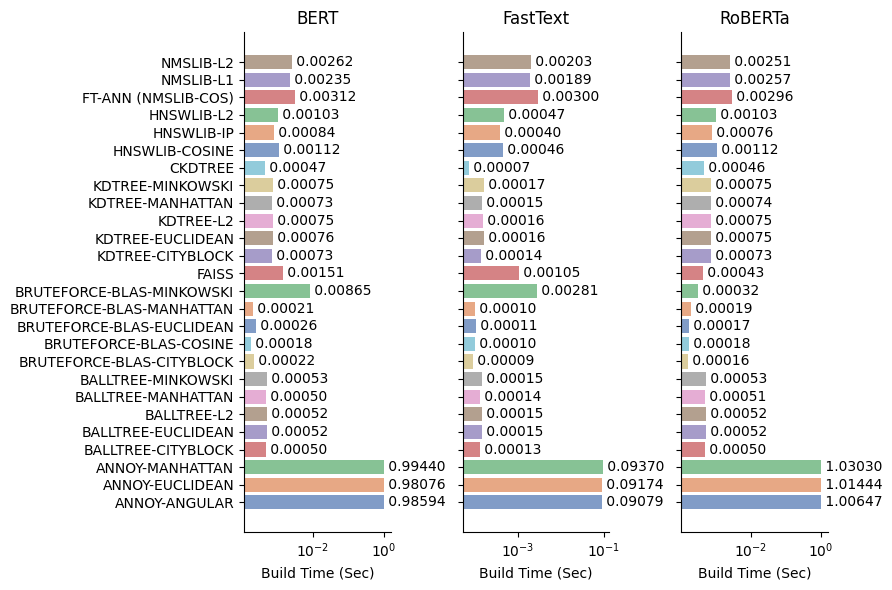}
\caption{Evaluation of Index Build Time (seconds) for In-Memory ANN Algorithms on the ATRDF 2023 Dataset. The Y-axis represents algorithm names, and the X-axis shows build time in log scale.}
\label{fig:atrdf_2023_ann_benchmark_build_time}
\end{figure}

\begin{figure}[!htb]
\centering
\includegraphics[width=0.5\textwidth]{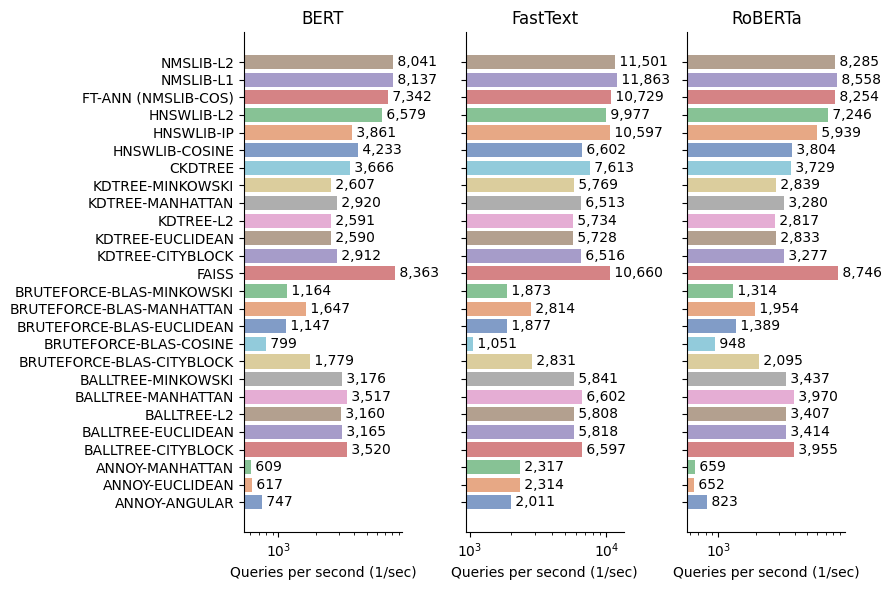}
\caption{Evaluation of QPS (queries per second) for In-Memory ANN Algorithms on the ATRDF 2023 Dataset. The Y-axis represents algorithm names, and the X-axis shows QPS in log scale.}
\label{fig:atrdf_2023_ann_benchmark_qps}
\end{figure}

Most ANN algorithms perform excellently on ATRDF 2023, so we chose to focus on comparing QPS and build time across different methods to evaluate their speed and scalability. The results are shown in Figures \ref{fig:atrdf_2023_ann_benchmark_build_time} and \ref{fig:atrdf_2023_ann_benchmark_qps}. FT-ANN using FastText embedding demonstrates a strong balance with high QPS (10,729.09) and reasonable build time (0.002997 seconds). In comparison, Nmslib-L1 and Nmslib-L2 outperform FT-ANN in terms of QPS, with values of 11,863.19 and 11,500.90, respectively, and slightly lower build times. Faiss and Hnswlib variants also offer high QPS with minimal build time. In contrast, BallTree variants achieve modest QPS with extremely low build times. Bruteforce-Blas models, despite their simplicity and extremely low build times, achieve lower QPS, making them less suitable for high-throughput applications but beneficial where build time is critical. Generally, we observe the consistency in the model's performance relative to each other under the different language model limitations.

\renewcommand{\arraystretch}{1.5}
\begin{table}[]
\resizebox{\columnwidth}{!}{%
\begin{tabular}{llll}
\hline
\textbf{Method}                      & \textbf{F1 Score} & \textbf{Precision} & \textbf{Recall} \\ \hline
FT-ANN                             & \textbf{0.9713}            & \textbf{0.9538}             & \textbf{0.9954}          \\ \hline
\cite{mac2018detecting}         & 0.9463            & 0.9464             & 0.9462          \\ \hline

\cite{demirel2023acum}         & 0.9274            & 0.9352             & 0.9333          \\ \hline
\cite{moradi2021auto}       & 0.89              & 0.8583             & 0.9243          \\ \hline
\cite{vartouni2018anomaly} & 0.8412            & 0.8029             & 0.8832          \\ \hline
\cite{moradi2019auto}       & 0.8196            & 0.8122             & 0.8273          \\ \hline
\end{tabular}%
}
\caption{Evaluation of CSIC 2010 dataset with various unsupervised methods.}
\label{tab:csic_2010_sota}
\end{table}

\renewcommand{\arraystretch}{1.5}
\begin{table*}[]
\resizebox{\textwidth}{!}{%
\begin{tabular}{llllllllllllllll}
\hline
 &
  \multicolumn{3}{c}{\textbf{F1 Score}} &
  \multicolumn{3}{c}{\textbf{Precision}} &
  \multicolumn{3}{c}{\textbf{Recall}} &
  \multicolumn{3}{c}{\textbf{Train Time (Sec)}} &
  \multicolumn{3}{c}{\textbf{Test Time (Sec)}} \\ \cline{2-16} 
\multirow{-2}{*}{} &
  \multicolumn{1}{l}{\textbf{BERT}} &
  \multicolumn{1}{l}{\textbf{FastText}} &
  \textbf{RoBERTa} &
  \multicolumn{1}{l}{\textbf{BERT}} &
  \multicolumn{1}{l}{\textbf{FastText}} &
  \textbf{RoBERTa} &
  \multicolumn{1}{l}{\textbf{BERT}} &
  \multicolumn{1}{l}{\textbf{FastText}} &
  \textbf{RoBERTa} &
  \multicolumn{1}{l}{\textbf{BERT}} &
  \multicolumn{1}{l}{\textbf{FastText}} &
  \textbf{RoBERTa} &
  \multicolumn{1}{l}{\textbf{BERT}} &
  \multicolumn{1}{l}{\textbf{FastText}} &
  \textbf{RoBERTa} \\ \hline
{\textbf{FT-ANN}} &
  \multicolumn{1}{l}{{1.0}} &
  \multicolumn{1}{l}{{\textbf{1.0}}} &
  {1.0} &
  \multicolumn{1}{l}{{1.0}} &
  \multicolumn{1}{l}{{\textbf{1.0}}} &
  {1.0} &
  \multicolumn{1}{l}{{1.0}} &
  \multicolumn{1}{l}{{\textbf{1.0}}} &
  {1.0} &
  \multicolumn{1}{l}{{0.0039}} &
  \multicolumn{1}{l}{{\textbf{0.0029}}} &
  {0.0042} &
  \multicolumn{1}{l}{{0.0012}} &
  \multicolumn{1}{l}{{\textbf{0.0004}}} &
  {0.0012} \\ \hline
\textbf{DeepSVDD} &
  \multicolumn{1}{l}{0.9953} &
  \multicolumn{1}{l}{0.996} &
  0.9944 &
  \multicolumn{1}{l}{0.9908} &
  \multicolumn{1}{l}{0.9921} &
  0.9889 &
  \multicolumn{1}{l}{1.0} &
  \multicolumn{1}{l}{1.0} &
  1.0 &
  \multicolumn{1}{l}{3.4848} &
  \multicolumn{1}{l}{3.186} &
  3.2811 &
  \multicolumn{1}{l}{0.0383} &
  \multicolumn{1}{l}{0.0332} &
  0.0476 \\ \hline
\textbf{AE} &
  \multicolumn{1}{l}{0.9951} &
  \multicolumn{1}{l}{0.9946} &
  0.9951 &
  \multicolumn{1}{l}{0.9903} &
  \multicolumn{1}{l}{0.9893} &
  0.9903 &
  \multicolumn{1}{l}{1.0} &
  \multicolumn{1}{l}{1.0} &
  1.0 &
  \multicolumn{1}{l}{5.043} &
  \multicolumn{1}{l}{6.0278} &
  4.9394 &
  \multicolumn{1}{l}{0.0421} &
  \multicolumn{1}{l}{0.0512} &
  0.0412 \\ \hline
\textbf{FB} &
  \multicolumn{1}{l}{0.9944} &
  \multicolumn{1}{l}{0.9946} &
  0.9944 &
  \multicolumn{1}{l}{0.9888} &
  \multicolumn{1}{l}{0.9893} &
  0.9888 &
  \multicolumn{1}{l}{1.0} &
  \multicolumn{1}{l}{1.0} &
  1.0 &
  \multicolumn{1}{l}{0.0207} &
  \multicolumn{1}{l}{0.0144} &
  0.021 &
  \multicolumn{1}{l}{0.0139} &
  \multicolumn{1}{l}{0.0055} &
  0.0146 \\ \hline
\textbf{IF} &
  \multicolumn{1}{l}{0.9944} &
  \multicolumn{1}{l}{0.9946} &
  0.9944 &
  \multicolumn{1}{l}{0.9888} &
  \multicolumn{1}{l}{0.9893} &
  0.9888 &
  \multicolumn{1}{l}{1.0} &
  \multicolumn{1}{l}{1.0} &
  1.0 &
  \multicolumn{1}{l}{0.2578} &
  \multicolumn{1}{l}{0.1915} &
  0.2597 &
  \multicolumn{1}{l}{0.0649} &
  \multicolumn{1}{l}{0.0307} &
  0.0651 \\ \hline
\textbf{KDE} &
  \multicolumn{1}{l}{0.9947} &
  \multicolumn{1}{l}{0.9946} &
  0.9947 &
  \multicolumn{1}{l}{0.9896} &
  \multicolumn{1}{l}{0.9893} &
  0.9896 &
  \multicolumn{1}{l}{1.0} &
  \multicolumn{1}{l}{1.0} &
  1.0 &
  \multicolumn{1}{l}{0.0043} &
  \multicolumn{1}{l}{0.001} &
  0.0043 &
  \multicolumn{1}{l}{0.0068} &
  \multicolumn{1}{l}{0.0005} &
  0.0069 \\ \hline
\textbf{LOF} &
  \multicolumn{1}{l}{0.9944} &
  \multicolumn{1}{l}{0.9946} &
  0.9944 &
  \multicolumn{1}{l}{0.9888} &
  \multicolumn{1}{l}{0.9893} &
  0.9888 &
  \multicolumn{1}{l}{1.0} &
  \multicolumn{1}{l}{1.0} &
  1.0 &
  \multicolumn{1}{l}{0.0015} &
  \multicolumn{1}{l}{0.0014} &
  0.0016 &
  \multicolumn{1}{l}{0.0015} &
  \multicolumn{1}{l}{0.0006} &
  0.0011 \\ \hline
\textbf{OCSVM} &
  \multicolumn{1}{l}{0.9939} &
  \multicolumn{1}{l}{0.9946} &
  0.9941 &
  \multicolumn{1}{l}{0.9879} &
  \multicolumn{1}{l}{0.9893} &
  0.9884 &
  \multicolumn{1}{l}{1.0} &
  \multicolumn{1}{l}{1.0} &
  1.0 &
  \multicolumn{1}{l}{0.0028} &
  \multicolumn{1}{l}{0.0011} &
  0.0028 &
  \multicolumn{1}{l}{0.0017} &
  \multicolumn{1}{l}{0.0002} &
  0.0016 \\ \hline
\textbf{PCA} &
  \multicolumn{1}{l}{0.9951} &
  \multicolumn{1}{l}{0.9946} &
  0.9951 &
  \multicolumn{1}{l}{0.9903} &
  \multicolumn{1}{l}{0.9893} &
  0.9903 &
  \multicolumn{1}{l}{1.0} &
  \multicolumn{1}{l}{1.0} &
  1.0 &
  \multicolumn{1}{l}{0.0132} &
  \multicolumn{1}{l}{0.0014} &
  0.014 &
  \multicolumn{1}{l}{0.0026} &
  \multicolumn{1}{l}{0.0002} &
  0.0028 \\ \hline
\textbf{HBOS} &
  \multicolumn{1}{l}{0.9938} &
  \multicolumn{1}{l}{0.9944} &
  0.9947 &
  \multicolumn{1}{l}{0.9878} &
  \multicolumn{1}{l}{0.9888} &
  0.9896 &
  \multicolumn{1}{l}{1.0} &
  \multicolumn{1}{l}{1.0} &
  1.0 &
  \multicolumn{1}{l}{0.4646} &
  \multicolumn{1}{l}{0.9126} &
  0.4715 &
  \multicolumn{1}{l}{0.005} &
  \multicolumn{1}{l}{0.0228} &
  0.0049 \\ \hline
\textbf{MCD} &
  \multicolumn{1}{l}{0.9939} &
  \multicolumn{1}{l}{0.9942} &
  0.9886 &
  \multicolumn{1}{l}{0.9879} &
  \multicolumn{1}{l}{0.9885} &
  0.9887 &
  \multicolumn{1}{l}{1.0} &
  \multicolumn{1}{l}{1.0} &
  0.9898 &
  \multicolumn{1}{l}{151.32} &
  \multicolumn{1}{l}{3.618} &
  174.05 &
  \multicolumn{1}{l}{0.2682} &
  \multicolumn{1}{l}{0.0005} &
  0.3775 \\ \hline
\textbf{CBLOF} &
  \multicolumn{1}{l}{0.9949} &
  \multicolumn{1}{l}{0.9896} &
  0.9957 &
  \multicolumn{1}{l}{0.99} &
  \multicolumn{1}{l}{0.9795} &
  0.9916 &
  \multicolumn{1}{l}{1.0} &
  \multicolumn{1}{l}{1.0} &
  1.0 &
  \multicolumn{1}{l}{0.0672} &
  \multicolumn{1}{l}{0.4408} &
  0.0609 &
  \multicolumn{1}{l}{0.002} &
  \multicolumn{1}{l}{0.0006} &
  0.002 \\ \hline
\textbf{LMDD} &
  \multicolumn{1}{l}{0.5641} &
  \multicolumn{1}{l}{0.292} &
  0.6261 &
  \multicolumn{1}{l}{1.0} &
  \multicolumn{1}{l}{0.8237} &
  1.0 &
  \multicolumn{1}{l}{0.3982} &
  \multicolumn{1}{l}{0.1825} &
  0.4628 &
  \multicolumn{1}{l}{0.3028} &
  \multicolumn{1}{l}{0.3387} &
  0.3208 &
  \multicolumn{1}{l}{1.2767} &
  \multicolumn{1}{l}{0.068} &
  1.3278 \\ \hline
\end{tabular}%
}
\caption{Performance evaluations of FT-ANN on the ATRDF 2023 dataset in comparison to baseline models for detecting outlying objects in multivariate data.}
\label{tab:ATRDF-2023-performance}
\end{table*}

\section{Discussion and Conclusions}
\label{sec:conclusions}
This paper suggests an innovative unsupervised few-shot anomaly detection framework that leverages a dedicated generic language model for API based on FastText embedding and uses ANN search in a Classification-by-retrieval approach. We demonstrated that API Injection attacks can be identified using only a few examples of normal API requests. Our framework not only accurately classifies the API traffic but also does it quickly and effectively with the support of incremental learning by the ANN, which is crucial for real-world systems. To the best of our knowledge, this is the first work to utilize a Classification-by-retrieval framework based on the generalized approach of FastText embedding combined with the approximate search to find anomalies in API traffic. 
Additionally, we presented a unique pre-processing technique to enhance input generalization and simplify API structure. This approach encompasses dividing input data into individual tokens and then constraining the vocabulary to a limited set of tokens. Consequently, the API structure becomes streamlined as the number of unique tokens diminishes, enabling input generalization and achieving high detection accuracy even with minimal examples per class. We presented several advanced anomaly detection models for this objective through two other widely adopted language models, BERT and RoBERTa. Our analysis also involved comparing various ANN search algorithms to demonstrate the framework's performance on the CSIC 2010 and ATRDF 2023 datasets. We compared our framework to the current SOTA unsupervised techniques and showed that it achieved the highest accuracy on the CSIC 2010 dataset.

While we observed that the cosine similarity implementation in Nmslib doesn't produce the highest F1 score among ANN algorithms, its straightforward implementation and seamless integration, and a good balance of QPS and build time make it a suitable choice for practical applications, particularly those like Amazon Elasticsearch Service that prioritize real-world compatibility. 

One notable limitation of this and similar studies arises from the scarcity of up-to-date and representative network traffic datasets. Many research efforts rely solely on the CSIC 2010 dataset, raising concerns about the reliability and applicability of such data, consequently impacting the quality and generalizability of the resulting models. While our evaluation focuses on measuring the impact of FT-ANN on the HTTP dataset, it is important to acknowledge that APIs exist in diverse forms, including REST, GraphQL, gRPC, and WebSockets, each catering to specific use cases and carrying common vulnerabilities. Our future work entails expanding the evaluation to encompass a wider range of API attacks and types. Additionally, we intend to assess the stability of the framework and its resource utilization under varying loads and volumes.

\section*{Funding}
This research received no specific grant from any funding agency in the public, commercial, or not-for-profit sectors.

\section*{CRediT authorship contribution statement}
Udi Aharon: Conceptualization, Data curation, Formal analysis, Investigation, Methodology, Project administration, Resources, Software, Validation, Visualization, Writing – original draft, Writing– review \& editing. Ran Dubin: Conceptualization, Methodology, Project administration, Supervision, Writing – review \& editing. Amit Dvir: Conceptualization, Investigation, Methodology, Writing – review \& editing. Chen Hajaj: Conceptualization, Data curation, Investigation, Methodology, Project administration, Writing – review \& editing.

\section*{Acknowledgment}
This work was supported by the Ariel Cyber Innovation Center in conjunction with the Israel National Cyber Directorate in the Prime Minister's Office.

\printcredits

\bibliographystyle{cas-model2-names}

\bibliography{cas-refs}

\newpage
\vskip3pt

\bio{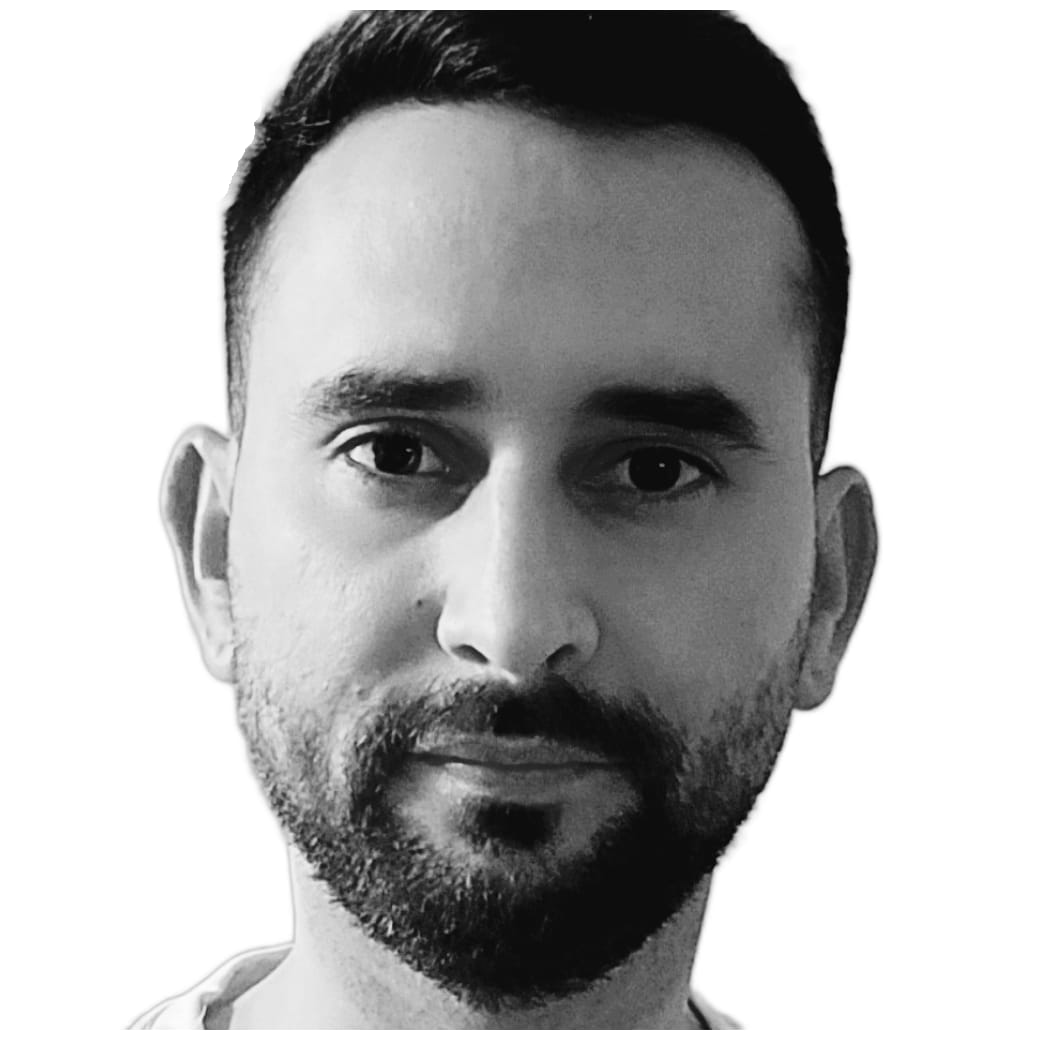}
Udi Aharon is currently pursuing a Ph.D. degree within the Department of Computer Science at Ariel University, Israel. His research activities span the fields of Machine Learning and Cybersecurity. Specifically, the primary focus of his work is on enhancing API security through the application of text-based models.
\endbio

\bio{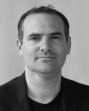}
Ran Dubin received his B.Sc., M.Sc., and Ph.D. degrees from Ben-Gurion University, Beer Sheva, Israel, all in communication systems engineering. He is currently a faculty member at the Computer Science Department, Ariel University, Israel. His research interests revolve around zero-trust cyber protection, malware disarms and reconstruction, encrypted network traffic detection, Deep Packet Inspection (DPI), bypassing AI, Natural Language Processing, and AI trust.
\endbio

\bio{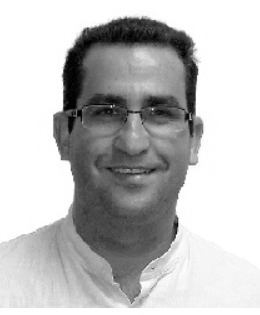}
Amit Dvir received his B.Sc., M.Sc., and Ph.D. degrees from Ben-Gurion University, Beer Sheva, Israel, all in communication systems engineering. He is currently a Faculty Member in the Department of Computer Science and the head of the Ariel Cyber Innovation Center, Ariel University, Israel. From 2011 to 2012, he was a Postdoctoral Fellow at the Laboratory of Cryptography and System Security, Budapest, Hungary. His research interests include enrichment data from encrypted traffic.
\endbio

\bio{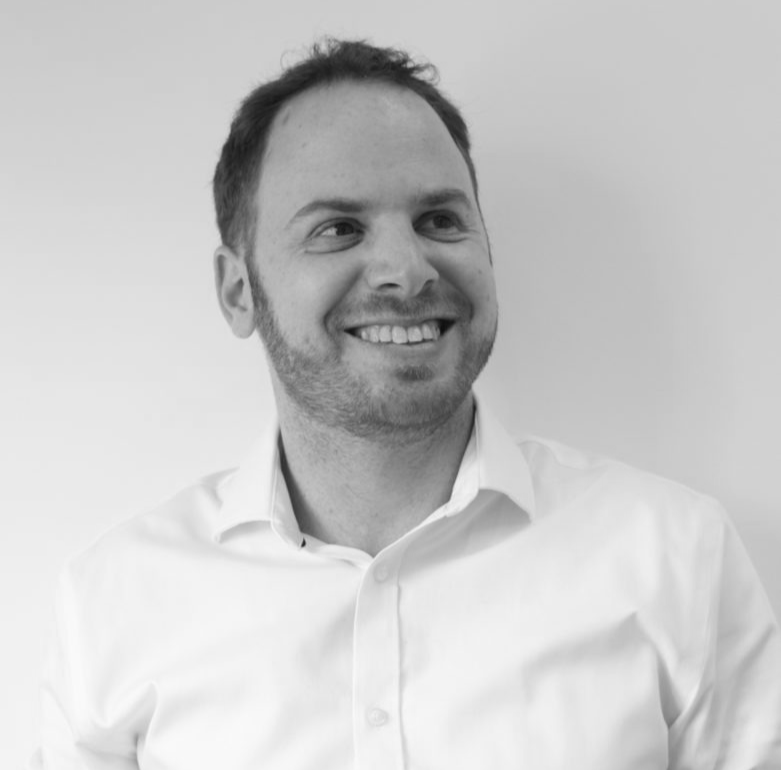}
Chen Hajaj holds Ph.D. (Computer Science), M.Sc. (Electrical Engineering), and B.Sc. (Computer Engineering) degrees, all from Bar-Ilan University. He is a faculty member in the Department of Industrial Engineering and Management, the head of the Data Science and Artificial Intelligence Research Center, and a member of the Ariel Cyber Innovation Center. From 2016 to 2018, Chen was a postdoctoral fellow at Vanderbilt University. Chen's research activities are in the areas of Machine Learning, Game Theory, and Cybersecurity. Specifically, the focuses of his work are on encrypted traffic classification, how to detect and robustify the weak spots of AI methods (adversarial artificial intelligence), and multimodal classification techniques.
\endbio

\end{document}